\documentclass[letterpaper]{article} % DO NOT CHANGE THIS
\usepackage{aaai2026}  % DO NOT CHANGE THIS
\usepackage{times}  % DO NOT CHANGE THIS
\usepackage{helvet}  % DO NOT CHANGE THIS
\usepackage{courier}  % DO NOT CHANGE THIS
\usepackage[hyphens]{url}  % DO NOT CHANGE THIS
\usepackage{graphicx} % DO NOT CHANGE THIS
\usepackage{natbib}  % DO NOT CHANGE THIS AND DO NOT ADD ANY OPTIONS TO IT
\usepackage{caption} % DO NOT CHANGE THIS AND DO NOT ADD ANY OPTIONS TO IT
\usepackage{amsmath}
\usepackage{multirow}
\usepackage{subfig}
\usepackage{enumitem}
\usepackage{soul}
\usepackage[ruled,linesnumbered]{algorithm2e}
\usepackage{amssymb}
\usepackage{booktabs}
\usepackage{tcolorbox}
\usepackage{dsfont}

\title{Adversarial Attack on Black-Box Multi-Agent by Adaptive Perturbation}
\author{
    Jianming Chen\textsuperscript{\rm 1,\rm 2,\rm 3,\rm 4},
    Yawen Wang\textsuperscript{\rm 1,\rm 2,\rm 3,\rm 4}\thanks{Corresponding authors.},
    Junjie Wang\textsuperscript{\rm 1,\rm 2,\rm 3,\rm 4}\footnotemark[1],
    Xiaofei Xie\textsuperscript{\rm 5},
    Yuanzhe Hu\textsuperscript{\rm 1,\rm 2,\rm 3,\rm 4},\\
    Qing Wang\textsuperscript{\rm 1,\rm 2,\rm 3,\rm 4},
    Fanjiang Xu\textsuperscript{\rm 1,\rm 2,\rm 3,\rm 4}\footnotemark[1]
}
\affiliations {
    \textsuperscript{\rm 1}
    Institute of Software Chinese Academy of Sciences, Beijing, China
    \\
    \textsuperscript{\rm 2}
    Science \& Technology on Integrated Information System Laboratory, Beijing, China
    \\
    \textsuperscript{\rm 3}
    State Key Laboratory of Complex System Modeling and Simulation Technology, Beijing, China
    \\
    \textsuperscript{\rm 4}
    University of Chinese Academy of Sciences, Beijing, China
    \\
    \textsuperscript{\rm 5}
    Singapore Management University, Singapore
    \\
    \{jianming2023, yawen2018, junjie\}@iscas.ac.cn,
    xfxie@smu.edu.sg,
    \{yuanzhe, wq, fanjiang\}@iscas.ac.cn
}

\usepackage{bibentry}
\begin{document}

\maketitle

\begin{abstract}
Evaluating security and reliability for multi-agent systems (MAS) is urgent as they become increasingly prevalent in various applications. As an evaluation technique, existing adversarial attack frameworks face certain limitations, e.g., impracticality due to the requirement of white-box information or high control authority, and a lack of stealthiness or effectiveness as they often target all agents or specific fixed agents. To address these issues, we propose AdapAM, a novel framework for adversarial attacks on black-box MAS. AdapAM incorporates two key components: (1) \textit{Adaptive Selection Policy} simultaneously selects the victim and determines the anticipated malicious action (the action would lead to the worst impact on MAS), balancing effectiveness and stealthiness. (2) \textit{Proxy-based Perturbation to Induce Malicious Action} utilizes generative adversarial imitation learning to approximate the target MAS, allowing AdapAM to generate perturbed observations using white-box information and thus induce victims to execute malicious action in black-box settings. We evaluate AdapAM across eight multi-agent environments and compare it with four state-of-the-art and commonly-used baselines. Results demonstrate that AdapAM achieves the best attack performance in different perturbation rates. Besides, AdapAM-generated perturbations are the least noisy and hardest to detect, emphasizing the stealthiness. 
\end{abstract}

% Uncomment the following to link to your code, datasets, an extended version or similar.
% You must keep this block between (not within) the abstract and the main body of the paper.
% \begin{links}
%     \link{Code}{https://aaai.org/example/code}
%     \link{Datasets}{https://aaai.org/example/datasets}
%     \link{Extended version}{https://aaai.org/example/extended-version}
% \end{links}

\section{Introduction}
Recent years have witnessed sensational advances in reinforcement learning (RL) across many prominent sequential decision-making problems \cite{RL1,RL2}. 
As these problems have grown in complexity, the field has transitioned from using primarily single-agent RL algorithms to multi-agent RL (MARL) algorithms, which are playing increasingly significant roles in various domains, e.g., unmanned aerial vehicles \cite{uav1,drones}, industrial robots \cite{robots2,multi-robot}, and auto-driving \cite{driving1,driving2}. 
These multi-agent systems (MAS) rely on multiple agents collaborating to achieve complex tasks, where each agent operates based on decentralized decision-making and shared information. However, the increasing deployment of such MAS has also made them attractive targets for adversarial attacks, raising concerns about their security and reliability in critical environments \cite{MAS-safe2}.

Adversarial attacks are techniques used to assess the security and reliability of artificial intelligence (AI) systems by deliberately introducing inputs that can mislead the system into making incorrect decisions \cite{usenix-adv_attack1,usenix-adv_attack2}.
For example, some imperceptible perturbations are added to the inputs of the attacked model to make that model produce the wrong output \cite{CCS-black-box1,CCS-black-box2}.
The adversarial attacks against machine learning (ML) systems have been extensively studied \cite{attack-ml1,attack-ml2}, including targeting single-agent systems \cite{attack-SA1,attack-SA2,attack-SA3}.
However, in MAS, the increasing interactions and dependencies between agents, as well as across time steps, introduce significant challenges that require further investigation to develop an effective adversarial attack framework \cite{EMAI}.

The current research on adversarial attacks against MAS usually relies on unrealistic assumptions about the problem setting. Many studies assume that attackers have high-level access to the victim MAS, such as requiring internal network outputs or other white-box information \cite{AMCA,Lin}, or even the action control authority of target agents \cite{attack-MA2,AMI}. 
However, such assumptions are impractical in practical applications. In reality, scenarios are more likely to involve the strict black-box setting, where attackers cannot directly access internal models or parameters of the system. Instead, they can only implement threats by perturbing the observations of the target agents \cite{black}.
This strict black-box setting is not only more practical but also presents greater challenges and is therefore more worthwhile to study.

Furthermore, existing adversarial attack methods for MAS often struggle with a trade-off between effectiveness and stealthiness. For example, some approaches \cite{MASafe,attack-MA1} add perturbations to all agents, which leads to poor stealthiness (in terms of the number of perturbed agents).
Some methods only consider specific members within the MAS as fixed adversaries \cite{AMI}, or fail to provide effective agent selection strategies \cite{AMCA}, making it difficult to achieve efficient attacks in the diverse state transitions of the multi-agent Markov Decision Process (MDP) \cite{MAMDP}.
Under constraints like limited attack budgets and stealthiness requirements, existing methods exhibit a lack of effectiveness and stealthiness.
Therefore, there is an urgent need for an attack method that can effectively and stealthily operate in strict black-box settings, thereby comprehensively assessing the robustness and safety of the MAS.

Drawing from the aforementioned issues, we propose \textbf{AdapAM}, a novel learning-based framework for \textbf{\textit{Adap}}tive adversarial \textbf{\textit{A}}ttacks on the black-box \textbf{\textit{M}}AS.
Our attack operates in a strict black-box setting, where only the observations and corresponding actions of the victim MAS can be accessed. 
Besides, only manipulating and perturbing the observations of agents is allowed, rather than actions. 
AdapAM adaptively selects the most important agent and determines the anticipated malicious action (achieved by specifically perturbing its observations), maximizing the effectiveness of the attack.
We define malicious action as the action we aim to induce the adversary to execute by perturbing its observation, which causes the worst impact on the victim MAS.
It avoids targeting multiple agents, thereby achieving both effectiveness and stealthiness in the attack.
The primary focus of this approach is to adaptively select the most important agent as the adversary and determine the corresponding malicious actions. 
We first design an adaptive selection policy, which learns to select the adversary and determine malicious action at each time step, based on the environment state.
The training of this policy is modeled as an RL process that optimizes the objective based on reducing the reward of the target agent.
Secondly, since determining the observations that can lead to malicious actions requires knowledge of the internal model (i.e., white-box setting), we conduct the perturbation generation through proxy agents while operating in a strict black-box setting. The proxy agents are trained in a generative adversarial imitation learning way with a mapping from observation to action that approximates the victim MAS. Therefore, the white-box information of proxy agents can be utilized to better produce perturbed observations.

We evaluate AdapAM in eight popular multi-agent environments and compare it with four state-of-the-art and commonly-used baselines.
The results indicate that, compared with the baselines, AdapAM achieves almost optimal attack performance against both normal MAS and robust MAS across all environments and various perturbation rate settings, including both reward and win rate metrics. 
Additionally, we evaluate the stealthiness of our AdapAM in two aspects: first, by calculating the distance between the observations before and after adding perturbations; second, by assessing the detection rate using existing attack detection methods to determine whether the victim MAS is under attack.
Experimental results show that the perturbations added by AdapAM are the smallest and the attacks it causes are the most difficult to detect, demonstrating that AdapAM has better stealthiness.

The main contributions of this work are as follows.
\begin{itemize}
    \item A novel black-box adversarial attack framework for MAS, learning an adaptive policy to select adversary and anticipated malicious action at each time step.
    \item The perturbations generated via proxy agents, where the proxy agents are the approximation of the victim MAS to provide white-box information and address the challenge of generating perturbations in a black-box setting.
    \item Experimental evaluation of attack performance, stealthiness of AdapAM in eight multi-agent tasks, which outperforms four state-of-the-art and commonly-used baselines and demonstrates the effects of adversary and malicious action selection.
\end{itemize}

\section{Related Work}
\subsection{Adversarial Attack}

Adversarial attack \cite{usenix-adv_attack1} is a type of attack targeting ML models, particularly deep neural network (DNN). By introducing perturbations to the input data, attackers aim to mislead the model into making incorrect predictions or classifications.
These perturbations are generated using either white-box \cite{JSMA,DeepFool} or black-box \cite{SimBA-black,Square-black} based methods and are usually undetectable to humans, but can have a significant impact on the outputs of models.
These findings have attracted a lot of attention since they suggest that ML models may lack robustness in some specific situations.

Regarding the adversarial attacks against agent systems, the studies \cite{attack-SA1} and \cite{attack-SA2} utilize gradient-based adversary attacks to generate adversarial perturbations of the state. However, they only mislead an agent to do a wrong action and may not lead to the minimal expected reward. In \cite{attack-SA3}, the adversary is modeled as a Markov decision process (MDP), and RL is employed to address it, which can work well in a low-dimensional state space. To extend this method to a high-dimensional one, the work \cite{attack-SA4} proposes a two-step attack framework for advising the worst-case action and generating perturbations.

Despite these advancements, the majority of existing work primarily concentrates on single-agent systems, often neglecting the critical aspect of selecting which agents should be designated as adversaries in multi-agent reinforcement learning (MARL) models. This consideration is vital, as the effectiveness of adversarial attacks can vary significantly depending on the specific agents targeted.

\subsection{Adversarial Attack on MAS}
\cite{Lin} attacks a default agent in the MAS by generating perturbed observations in a JSMA-based algorithm \cite{JSMA}, which is a white-box approach. 
\cite{MASafe} and \cite{attack-MA1} consider all agents in the MAS as victims of the attack. The attack has superior performance due to the large number of agents affected, but such attacks are extremely easy to detect.
\cite{AMCA} proposes an optimization algorithm optimized with a joint action-value function to select key agents as victims for the attack. However, the value function is unavailable in strict black-box situations. In addition, the current environmental state is not considered in the selection.
Considering only attacking a fixed victim or not selecting a victim based on the state of the environment leads to ineffective attacks.

Attacking the actions of the agent in MAS is another scenario. \cite{attack-MA2} shows that in a two-agent competitive environment, the actions of one agent can be manipulated to fool the other.
\cite{AMI} controls a fixed agent in a cooperative MAS, learning the action that is most harmful to the other agents.
However, it is a strong assumption to have action manipulation permissions for the target agent, which is a rare case in practice.

\section{Threat Model and Problem Statement}
\label{sec:Threat_Problem}

\subsection{Threat Model}
\label{sec:Threat}
In this paper, we propose an adaptive adversarial attack targeting black-box MAS.

\textbf{Definition 1.} A MAS consists of a set of agents, denoted as \( M = \{ m_1, m_2, \ldots, m_n \} \), where \( |M| = n \geq 2 \). Each agent $m_i$ has the independent local observation space \( O_i \) and the action space \( A_i \). MAS typically maintains internal collaboration mechanisms to collectively accomplish a task (e.g., to counter another MAS or to achieve some goals).

\textbf{Attacker's Objective:} 
Previous work \cite{EMAI, AMCA} has shown that attacking only a small number of the most critical agents can significantly degrade the performance of the victim MAS.
Therefore, if it is possible to select an adversary at each time step adaptively, it can avoid perturbing a large number of agents to ensure stealthiness. On the other hand, the selected adversary needs to have a sufficient influence on the victim MAS to ensure the effectiveness of the attack.
Based on this, the attacker's objective is to adaptively select an agent as the target adversary at each time step, using perturbations to induce it to execute specific malicious actions to degrade the performance of the MAS.
 
\textbf{Attacker's Capability:} The attacker possesses control over the environment, which is a common situation.
Specifically, during the training phase, the attacker can fully control the actions of the adversary. For instance, in real-world multi-agent autonomous driving, a driver can seize control to override the outputs of the autonomous agent. In the simulation environment, we can achieve this by tampering with the action signals received by the environment.
In the deployment phase of the attack, the attacker can only manipulate the state observation provided to the adversary agent, aiming to mislead its policy through perturbations. 
It is a strict setting, which can be achieved by adding a patch as a perturbation to the agent camera \cite{patch}.

We assume that the victim MAS policy is fixed, i.e., the parameters in the deployment phase are frozen \cite{multi-robot, drones}.
Finally, the attack we study adheres to a strict black-box setting, where control of actions for the selected adversary is achieved through environmental perturbations, without access to the algorithm used for training or constructing agents and their network architecture.

\begin{figure}[t]
    \centering
    \includegraphics[width=0.99\columnwidth]{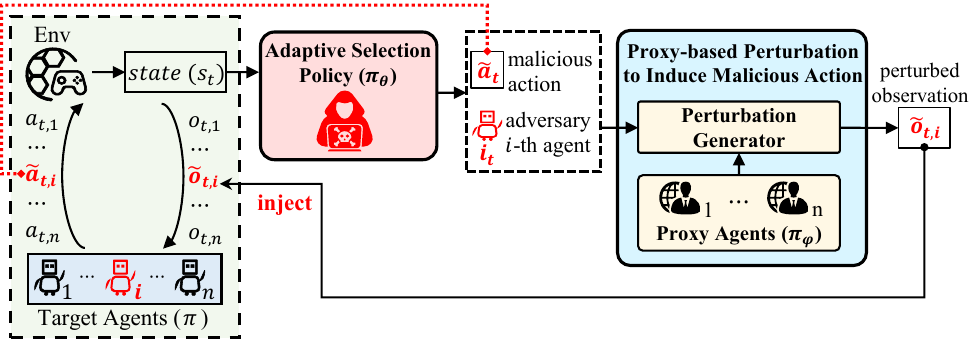}
    \caption{
    The overview of our proposed AdapAM.
    }
    \label{fig:overview}
\end{figure}

\subsection{Problem Statement}
\label{sec:problem}
Formally, the multi-agent Markov Decision Process (MDP) \cite{MDP} is defined as follows:
\begin{equation}
    \label{equation:mas-mdp}
    G = (n, \mathcal{S}, \{\mathcal{A}_i\}, \{\mathcal{O}_i\}, \pi, T, R),
\end{equation}
where $n$ represents the number of agents in the system, indicating the total count of decision-makers involved.
$\mathcal{S}$ denotes the global state space, which encompasses all possible states $s$ that describe the current configuration of all agents and their environment.
$\mathcal{A}_i$ denotes the set of actions that the $i$-th agent can choose from, while the joint action space of MAS can be represented as the Cartesian product of all individual action spaces.
The observation space $\mathcal{O}_i$ refers to the set of information that the $i$-th agent can perceive, which is typically a subset of the global state, reflecting the partial information available to the agents.
We consider a problem setting where a joint policy $\pi = \{\pi_1, ..., \pi_n\}$ has been well trained for $n$ agents in the victim MAS. At each time-step $t$ in an episode, the $i$-th agent obtains its local observation $o_{t, i}$ derived from the global environment state $s_t$ based on the observation function.
The policy $\pi_i(a_{t,i} | o_{t, i})$ of the $i$-th agent determines the action $a_{t,i}$ solely based on its local observation $o_{t, i}$.
The joint action $a_t = \{a_{t, 1}, ..., a_{t, n}\}$ transitions the system to the next state $s_{t+1}$ according to the state transition function $T(s_{t+1} | s_t, a_t)$.
Thereby, a global reward $r_t$ is obtained according to the reward function $R(s_t, a_t, s_{t+1})$.

We further decompose the problem into two subproblems, as follows:

\textbf{(1) Adaptive Selection Policy}

The attacker’s policy is defined as the following MDP:
\begin{equation}
    \label{equation:attack-mdp}
    G^\alpha = (\mathcal{S}, \mathcal{A}^\alpha, T, R^\alpha).
\end{equation}

Retain the same state space $\mathcal{S}$ and state transition function $T$ as in Equation \ref{equation:mas-mdp}. However, the attacker's policy needs to adaptively select the agent to target as an adversary and the corresponding malicious action based on the state.
$\mathcal{A}^\alpha$ denotes the action space of the policy, which outputs the attack action $\tilde{a}_i$. $\tilde{a}_i$ represents the i-th agent (serves as adversary) and its corresponding malicious action $\tilde{a}$ (sampled from $\mathcal{A}_i$).
In addition, the attacker's reward $R^\alpha$ aims to degrade the performance of the victim MAS.

\textbf{(2) Perturbation to Induce Malicious Action}

When we obtain $\tilde{a}_i$, we do not directly control the adversary's actions to execute the attack. Instead, we generate perturbed observation for the adversary to induce it to output the malicious actions we anticipate. Therefore, we need a perturbed observation $\tilde{o}_i$, as follows:
\begin{equation}
\label{equation:perturbed}
    \begin{aligned}
        & \tilde{o}_i = o_i + \delta, \quad \\
        & \tilde{o}_i = \arg\max_{\substack{||\delta|| \leq \epsilon}} \mathds{1}(\pi_i(o_i + \delta) = \tilde{a}_i),
    \end{aligned}
\end{equation}
where $\delta$ is the adversarial perturbation to be added, $\epsilon$ represents the scale of the perturbation, and $\mathds{1}(\cdot)$ is the indicator function (which is true when $\pi_i$ outputs the malicious action $\tilde{a}_i$ based on the perturbed observation $o_i + \delta$).
We need to find a perturbation that satisfies both Equation \ref{equation:perturbed}.

\section{Approach}
\label{sec:approach}

We propose AdapAM, a framework to adaptively select the adversary and determine malicious action on the target MAS. The adversary represents the most critical agent in the victim MAS, while the malicious action means the action induced by perturbing the adversary's observation to cause the worst impact on the target MAS.
AdapAM utilizes white-box information from proxy agents to generate a perturbed observation, which is then injected into the observation of adversary.
As shown in Figure \ref{fig:overview}, AdapAM primarily consists of two modules: \textbf{Adaptive Selection Policy} and \textbf{Proxy-based Perturbation to Induce Malicious Action}. 

The target MAS comprises $n$ agents interacting with the environment. As previously discussed, each agent plays a distinct role, whose actions have varying degrees of impact. To maximize the effectiveness of an attack under a limited attack budget,
it is necessary to select the most important agent as the adversary and the most malicious action based on the environmental state. Therefore, we design an adaptive selection policy to select the adversary agent $i_t$ and a specific malicious action $\tilde{a}_t$ desired to be performed by $i_t$.

To mislead $i_t$ into executing the action $\tilde{a}_t$ in a black-box setting, the proxy agents are required to be trained in advance first. The policies of proxy agents approximate those of target agents, ensuring transferability between the two. Utilizing the gradient information provided by the proxy model, we can employ the white-box method to generate a perturbed observation $\tilde{o}_{t,i}$ corresponding to the desired malicious action $\tilde{a}_t$. The $\tilde{o}_{t,i}$ is then injected into the observation of the adversary agent, effectively inducing it to execute $\tilde{a}_t$.

\begin{figure}[t]
    \centering
    \includegraphics[width=0.92\columnwidth]{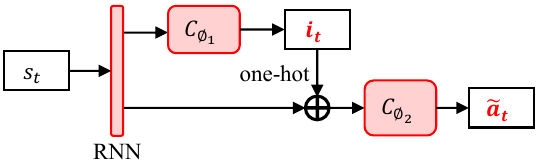}
    \caption{
    The architecture of the Adaptive Selection Policy.
    }
    \label{fig:attack_architecture}
\end{figure}

\subsection{Adaptive Selection Policy}

We design a learning-based adaptive selection policy, which aims to adaptively select the adversary agent and determine its anticipated malicious action according to the environment state, thus significantly degrading the performance of the entire MAS with a limited attack budget.

\begin{algorithm}[ht]
\caption{Adaptive Selection Policy Learning}
\label{alg:attack_policy}
\KwIn{Initial network parameters $\phi$, $\theta_1$, $\theta_{2}$, $\bar{\theta}_1$, $\bar{\theta}_2$}
\KwOut{Updated parameters $\phi$, $\theta_{1}$, $\theta_{2}$}
\textbf{Initialization:} episode buffer $\mathcal{D} \gets \emptyset$, weight parameter $\mu=0.005$   \\
\ForEach{iteration}{
    \ForEach{environment step}{
        Sample $(i_t, \tilde{a}_t) \sim \pi_{\phi}(i_t, \tilde{a}_t | s_t)$    \\
        Generate perturbation $\tilde{o}_{t,i}$, which induces adversary agent $i_t$ to take action $\tilde{a}_t$   \\
        Inject perturbation $o_{t,i} \gets \tilde{o}_{t,i}$   \\
        Sample $a_{t} \sim \pi(a_{t} | o_{t})$    \\
        Sample next state $s_{t+1} \sim T(s_{t+1} | s_t, a_t)$    \\
        Get attack reward $r^{a}_t = -R(s_t, a_t, s_{t+1})$    \\
        Save data $\mathcal{D} \gets \mathcal{D} \cup \{(s_t, i_t, \tilde{a}_t, r^{a}_t, s_{t+1})\}$    \\
    }
    \ForEach{gradient step}{
        \For{$k \in \{1, 2\}$}{
            Update critic: $\theta_k \gets \theta_k - \hat{\nabla}_{\theta_k} J_Q(\theta_k)$    \\
        }
        Update policy: $\phi \gets \phi - \hat{\nabla}_\phi J_\pi(\phi)$    \\
        \For{$k \in \{1, 2\}$}{
            Update target critic: $\bar{\theta}_i \gets \mu \theta_k + (1 - \mu) \bar{\theta}_i$    \\
        }
    }
}
\end{algorithm}

The adaptive selection policy $\pi_{\phi}(i_t, \tilde{a}_t | s_t)$, parameterized by $\phi$, learns to determine which adversary agent $i_t$ to attack and the malicious action $\tilde{a}_t$ it intends the adversary to take, at the time-step $t$ and given the current environmental state $s_t$. 
If we directly learn $i_t$ and $\tilde{a}_t$ simultaneously, the search space is large, making convergence difficult.
Therefore, technically, $\pi_{\phi}(i_t, \tilde{a}_t | s_t)$ actually includes two classifiers: $C_{\phi_1}(i_t | s_t)$ for selecting $i_t$ and $C_{\phi_2}(\tilde{a}_t | s_t, i_t)$ for selecting $\tilde{a}_t$. After $C_{\phi_1}(i_t | s_t)$ output $i_t$, we concatenate $s_t$ with the one-hot vector of $i_t$, and then input them together into the $C_{\phi_2}(\tilde{a}_t | s_t, i_t)$ for selecting $\tilde{a}_t$.
As shown in Figure \ref{fig:attack_architecture}, it is a hierarchical design to help reduce the search space

Besides, we design the learning process of $\pi_{\phi}(i_t, \tilde{a}_t | s_t)$ based on SAC \cite{SAC1}, as shown in Algorithm \ref{alg:attack_policy}.
The process alternates between collecting experience buffers $\mathcal{D}$ from the environment with the current policy and updating the network parameters using the stochastic gradients from batches sampled from $\mathcal{D}$.

Lines 2 to 11 of the Algorithm \ref{alg:attack_policy} describe the procedure of collecting buffers.
At each time-step, $i_t$ and $\tilde{a}_t$ are sampled from $\pi_{\phi}(i_t, \tilde{a}_t | s_t)$.
Then AdapAM generates the corresponding perturbed observation $\tilde{o}_{t,i}$ based on $i_t$ and $\tilde{a}_t$, using the module of \textit{Proxy-based Perturbation to Induce Malicious Action}, which will be presented later.
$\tilde{o}_{t,i}$ will then be injected into the observation of the target agent $i_t$, and the target MAS decides the actions based on the injected observations.
After calculating the next state according to the state transition function and corresponding attack reward $r^{a}_t$, the episode data are stored into $\mathcal{D}$, where $r^{a}_t=-R(s_t, a_t, s_{t+1})$.

Lines 12 to 20 of the Algorithm \ref{alg:attack_policy} describe the procedure of updating network parameters. The Q-function (i.e., critic) parameters can be trained to minimize the following soft Bellman residual \cite{SAC}:
\begin{equation}
\label{J_Q}
    \begin{aligned}
    & y = r^{a}_t + \gamma (\bar{\theta}(s_{t+1}, i_{t+1}, \tilde{a}_{t+1})   \\
    & - \alpha\log \pi_{\phi}(i_{t+1}, \tilde{a}_{t+1} | s_{t+1})),    \\
    & J_Q(\theta) = \mathds{E}_{(s_t, i_t, \tilde{a}_t) \sim \mathcal{D}} [ \frac{1}{2} ( Q_\theta(s_t, i_t, \tilde{a}_{t}) - y)^2 ],
    \end{aligned}
\end{equation}
where $\alpha$ is a weighting parameter to balance exploration and learning.
The policy parameters can be updated by minimizing the following expected KL-divergence:
\begin{equation}
\label{J_p}
    \begin{aligned}
    & J_\pi(\phi) = \mathds{E}_{s_t \sim \mathcal{D}}[\mathds{E}_{(i_t, \tilde{a}_{t}) \sim \pi_\phi}[\alpha\log(\pi_\phi(i_t, \tilde{a}_{t} | s_t))   \\
    & - Q_\theta(s_t, i_t, \tilde{a}_{t})]].
    \end{aligned}
\end{equation}

In line 14 and line 16 of Algorithm \ref{alg:attack_policy}, the $\hat{\nabla}_{\theta_k} J_Q(\theta_k)$ and $\hat{\nabla}_\phi J_\pi(\phi)$ represent the gradients of $J_Q(\theta)$ and $J_\pi(\phi)$, which are used to update the critic and policy networks, respectively.
Besides, in line 14 and line 18, we use 2 critic networks and 2 target critic networks to calculate Q-values following the setting in SAC \cite{SAC}, which have been shown to stabilize and boost training.

\begin{figure}[t]
    \centering
    \includegraphics[width=0.99\columnwidth]{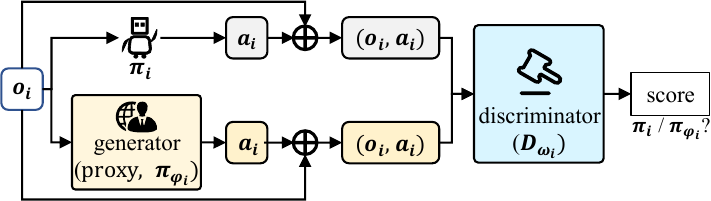}
    \caption{
    The setting of the training proxy agent.
    }
    \label{fig:proxy}
\end{figure}

\subsection{Proxy-based Perturbation to Induce Malicious Action}
\label{sec:Perturbation-Proxy}
To efficiently generate perturbed observation that induces adversaries to perform specific malicious actions in the black-box MAS, we train proxy agents to provide white-box information for perturbation generation.

\subsubsection{Training the Proxy Agents}
The policy of the $i$-th target agent $\pi_i(a_{i} | o_{i})$, represents the mapping $o_{i}\to a_{i}$, from observation to action.
If internal information of the mapping can be obtained (such as the intermediate layers' outputs and network gradients), it would be easy to generate perturbations that induce specific actions. However, this information cannot be directly accessed in a black-box setting
To this end, a proxy agent is needed, which learns a policy (denoted as $\pi_{\psi_i}$) to approximate this mapping, to provide a pipeline for generating perturbed observation from malicious action:
\begin{equation}
\label{eq-proxy}
    \pi_{\psi_i}(a_{i} | o_{i}) \approx \pi_i(a_{i} | o_{i}).
\end{equation}

We utilize Multi-Agent Generative Adversarial Imitation Learning (MAGAIL) \cite{MAGAIL} to train proxy agents $\pi_{\psi_i}$. The proxy agents imitate the behavior of target agents by learning their policy through adversarial imitation and serve as surrogates for generating perturbations.

As shown in Figure \ref{fig:proxy}, for each $i$-th agent, we have a discriminator (denoted as $D_{\omega_i}$) and a generator (denoted as $\pi_{\psi_i}$), i.e., proxy policy.
The $D_{\omega_i}$ maps observation-action pairs to the score, which is optimized to discriminate the action sampled by the $i$-th target agent (with policy $\pi_i$) from the action sampled by proxy policy $\pi_{\psi_i}$.

\subsubsection{Perturbation Generator}
After the proxy agents are trained, the internal information can be used to generate the perturbed observation corresponding to the malicious action.
Given the malicious action $\tilde{a}_{t}$ at time-step $t$, the goal of the perturbation generator is to find $\tilde{o}_{t,i}$ such that $argmax \pi_i(\tilde{o}_{t,i})=\tilde{a}_{t}$.

By utilizing the proxy agents, we leverage the white-boxed C\&W attack technique \cite{CW} to generate the perturbed observation.
This method can misguide the target agent with a 100\% success rate, which can increase the success rate of the attack.
The $\tilde{o}_{t,i}$ can be found by minimizing the following function:
\begin{equation}
\label{eq-CW}
    \begin{aligned}
    ||{o}_{t,i} - \tilde{o}_{t,i}|| + f(\tilde{o}_{t,i}).
    \end{aligned}
\end{equation}

The first term in Equation \ref{eq-CW} represents the distance between the perturbed observation and the original observation, i.e., the magnitude of the perturbation.
We use Euclidean distance \cite{L2} to ensure that small changes in each pixel point are calculated.
The second term represents the objective function of ensuring that $\pi_i(\tilde{o}_{t,i})=\tilde{a}_{t}$, i.e., it measures whether the input $\tilde{o}_{t,i}$ leads the policy $\pi_i$ to output the target action $\tilde{a}_{t}$.
Suggested by \cite{CW}, we define $f(\tilde{o}_{t,i})$ as:
\begin{equation}
\label{eq-CW-f}
    \begin{aligned}
    f(\tilde{o}_{t,i}) = \max (\max \{Z(a|\tilde{o}_{t,i}, a\neq \tilde{a}_{t})\} - Z(\tilde{a}_{t}|\tilde{o}_{t,i}), 0),
    \end{aligned}
\end{equation}
where $Z(a | \tilde{o}_{t,i})$ denotes the output of the second-to-last network layer (i.e., the logits output), with $a$ as the target action.

\section{Experiment}
\subsection{Experimental Setup}
\label{sec:exp_setup}

\begin{table*}[t]
    \centering
    \resizebox{\textwidth}{!}{
    \begin{tabular}{c||c||c|c|c|c|c|c||c|c|c|c|c|c}
    \toprule
    \multirow{2}{*}{\textbf{Env}}    & \multirow{2}{*}{\textbf{Metric}}         & \multicolumn{6}{c||}{\textbf{Normal Target MAS}}                & \multicolumn{6}{c}{\textbf{Robust Target MAS}} \\ 
                                     &                                          & Origin & AdapAM (ours) & MASafe & AMCA & AMI & Lin     & Origin & AdapAM (ours) & MASafe & AMCA & AMI & Lin \\
    \midrule
    \multirow{2}{*}{SMAC-1c3s5z}
    & Reward        & 20        & \textbf{9.02}    & 9.36             & 10.34            & 12.27   & 13.68       & 19.06     & \textbf{9.12}    & 10.79   & 11.02   & 12.59  & 14.80  \\
    & Win Rate      & 100\%     & \textbf{0\%}     & \textbf{0\%}     & \textbf{0\%}     & 17\%    & 25\%        & 91\%    & \textbf{0\%}       & \textbf{0\%}    & \textbf{0\%}    & 19\%   & 29\% \\
    \midrule
    \multirow{2}{*}{SMAC-8m}
    & Reward        & 20        & \textbf{9.57}    & 11.08            & 12.36           & 13.30    & 14.56       & 19.28     & \textbf{9.80}     & 11.63  & 12.05  & 13.16  & 15.08  \\
    & Win Rate      & 100\%     & \textbf{0\%}     & 6\%              & 17\%            & 25\%     & 32\%        & 94\%    & \textbf{0\%}        & 10\%    & 14\%   & 22\%   & 37\%   \\
    \midrule
    \multirow{2}{*}{SMAC-bane\_vs\_bane}
    & Reward        & 20        & 13.21   & \textbf{12.25}            & 14.63           & 15.52    & 16.02      & 18.79    & 13.73  & \textbf{13.05}       & 14.82      & 15.22          & 15.70  \\
    & Win Rate      & 100\%     & 22\%    & \textbf{16\%}             & 30\%            & 36\%     & 43\%      & 90\%     & 24\%   & \textbf{21\%}        & 31\%       & 33\%      & 38\%  \\
    \midrule
    \multirow{2}{*}{SMAC-27m\_vs\_30m}
    & Reward        & 19.24     & 13.87   & \textbf{13.19}            & 14.94           & 15.68    & 16.78      & 18.55    & 14.33  & \textbf{13.36}       & 15.21     & 15.59          & 16.24  \\
    & Win Rate      & 94\%      & 28\%    & \textbf{23\%}             & 31\%            & 39\%     & 47\%      & 84\%     & 29\%   & \textbf{25\%}        & 33\%      & 37\%      & 45\% \\
    \midrule
    \multirow{2}{*}{GF-counterattack}
    & Reward        & 5.21      & \textbf{0.56}    & 0.59             & 0.69            & 0.95    & 1.08   & 4.89     & \textbf{0.49}  & 0.78   & 0.76   & 1.08   & 1.27   \\
    & Win Rate      & 100\%     & \textbf{0\%}     & \textbf{0\%}     & 2\%             & 5\%     & 6\%    & 91\%     & \textbf{0\%}   & 3\%    & 3\%    & 6\%    & 13\%   \\
    \midrule
    \multirow{2}{*}{GF-3\_vs\_1}
    & Reward        & 5.43      & \textbf{0.99}    & 1.07             & 1.21            & 1.37    & 1.67        & 5.06      & \textbf{0.87}  & 1.19   & 1.24   & 1.36   & 1.75   \\
    & Win Rate      & 100\%     & \textbf{0\%}     & \textbf{0\% }    & 12\%             & 16\%     & 21\%        & 98\%    & \textbf{0\%}   & 10\%    & 12\%    & 16\%    & 23\%   \\
    \midrule
    MPE-spread
    & Reward        & -546.99   & \textbf{-1059.49}  & -1056.90  & -1018.49  & -1006.24  & -968.30       & -596.13       & \textbf{-1024.21} & -1006.11 & -971.57 & -972.15 & -953.28 \\
    \midrule
    MPE-reference
    & Reward        & -9.72     & \textbf{-38.80}    & -37.71    & -36.61    & -36.82    & -35.08               & -11.06        & \textbf{-36.49}   & -34.95  & -34.42  & -34.26  & -33.74 \\
    \bottomrule
    \end{tabular}
    }
    \caption{
    Attack performance of different environments and MAS, measured by the reward or win rate (the lower, the better).
    }
    \label{tab:attack_performance}
\end{table*}

\begin{figure*}[t]
    \centering
    \includegraphics[width=0.99\textwidth]{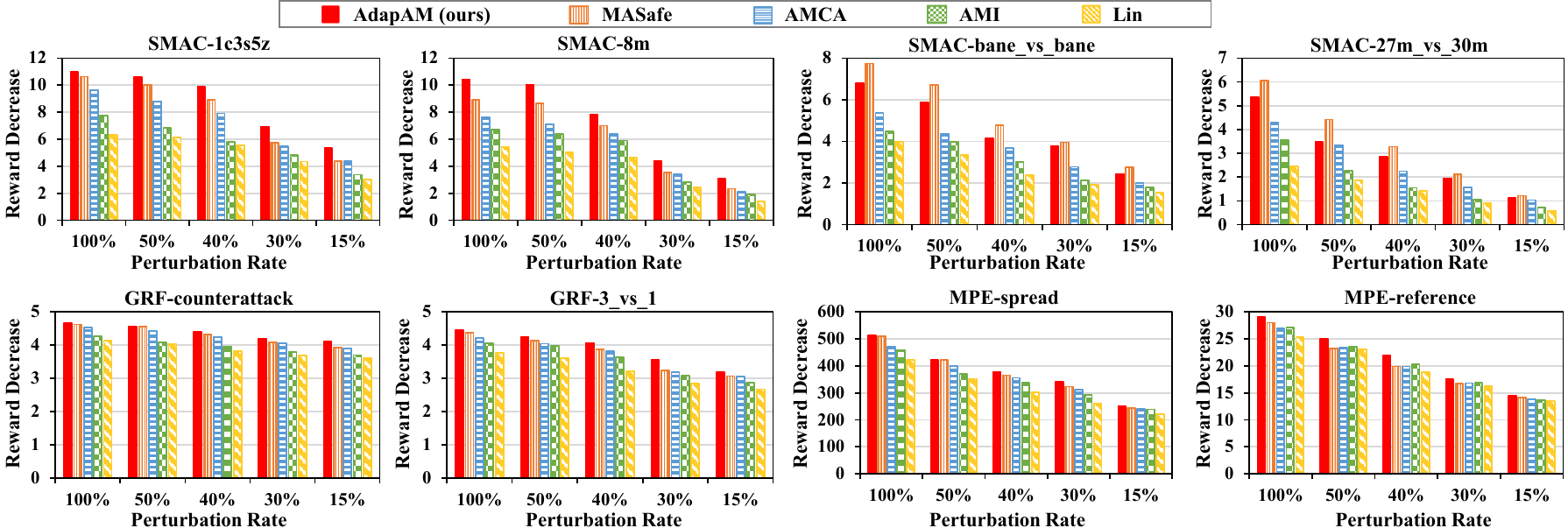}
    \caption{
    The decrease in reward after being attacked at different perturbation rates, on the \textit{Normal Target MAS}. 
    }
    \label{fig:attack_rate-reward-normal}
\end{figure*}

% \subsubsection{Multi-Agent Environments}
Our experiments are conducted on three popular multi-agent benchmarks with different characteristics, selecting two to three environments from each benchmark as follows.
% We introduce more experimental details in Appendix A.

\textbf{StarCraft Multi-Agent Challenge (SMAC).}
SMAC \cite{SMAC} simulates battle scenarios in which a team of controlled agents must destroy the built-in enemy team.
We consider two environments in SMAC, which vary in the number and types of units controlled by agents.

\textbf{Google Research Football (GF).}
GF \cite{GRF} provides the scenarios of controlling a team to play football against the built-in team.
We choose two environments in GF, which vary in the number of players and the tactics.

\textbf{Multi-Agent Particle Environments (MPE).}
MPE \cite{MPE} consists of navigation tasks, where agents need to control particles to reach the target landmarks.
We study two of these tasks, which mainly differ in whether communication is required between agents.

% \subsubsection{Baseline Approaches}
We implement four state-of-the-art and popular baseline approaches for adversarial attacks in each multi-agent environment.

\textbf{MASafe} \cite{MASafe}: applying the random perturbations to the observations of all agents. 

\textbf{AMCA} \cite{AMCA}: identifying important agents with a differential evolutionary algorithm and generating state perturbations after learning malicious actions.
                                                     
\textbf{AMI} \cite{AMI}: directly controlling the default agent, learning attack actions based on mutual information. It needs access to control the target agent.

\textbf{Lin} \cite{Lin}: generating observation perturbations corresponding to malicious actions against the default and fixed agent.

\subsection{Attack Performance}
\label{sec:attack_performance}

We evaluate the attack performance of our method in terms of its ability to degrade the overall functionality of a multi-agent system (MAS). Specifically, the effectiveness of the attack is measured by the reduction in the \textit{Reward} and \textit{Win Rate} of the target system across multiple benchmark environments. 
In addition, our attacks target two types of MAS: \textit{Normal Target MAS}, which represents common MAS trained using basic MARL methods (using QMIX \cite{QMIX}), and \textit{Robust Target MAS}, which represents the MAS with enhanced resistance to attacks, trained using a robustness-enhancing method called ROMANCE  \cite{ROMANCE}.
By introducing attackers, ROMANCE enables the policy to encounter diverse adversarial attacks as an auxiliary during adversarial training, so that it is trained to be highly robust under various perturbations.

Table~\ref{tab:attack_performance} compares the experimental results of the two target MAS in terms of \textit{Reward} and \textit{Win Rate}, in case of the \textit{origin} (non-attacked) performance and the performance under different attack methods. 
Notably, win or loss is not defined in the MPE-spread and MPE-reference environments, so attack performance is measured only by reward.

Figure \ref{fig:attack_rate-reward-normal} compares the attack performance in the case of different perturbation rates on normal target MAS, where the attack performance is represented by the decrease in \textit{reward} and \textit{win rate}, respectively.
The higher the decrease, the better the attack performance.
The perturbation rate represents the percentage of time-steps to perform the attack in each episode.
The results in Table \ref{tab:attack_performance} are for cases where the perturbation rate is 100\%.
% More results for different perturbation rates are shown in Appendix C.

\noindent\textbf{Results.} 
In all environments, our proposed AdapAM significantly reduces both the reward and win rate. In the environments where the number of agents is less than 10, AdapAM effectively degrades the MAS's ability to achieve its objectives (i.e., the Win Rate drops to $0\%$) and outperforms all baselines.
Only when attacking in the \textit{SMAC-bane\_vs\_bane} and \textit{SMAC-27m\_vs\_30m} environments, AdapAM does not achieve optimal performance, being slightly weaker than MASafe. 
While MASafe perturbs all agents, we only choose one agent to attack. Even with a significant difference in the number of perturbed agents, AdapAM does not fall far behind MASafe, which further demonstrates the importance of selecting the target agent and malicious actions.

Besides, when attacking the Robust Target MAS, AdapAM achieves a more significant attack effect in all cases. It shows the capability of AdapAM to maintain superior attack performance against robust MAS.
Similarly, as shown in Figure \ref{fig:attack_rate-reward-normal}, AdapAM almost always outperforms baseline methods under different perturbation rates, except in environments (i.e., \textit{SMAC-bane\_vs\_bane} and \textit{SMAC-27m\_vs\_30m}) with a large number of agents where it is outperformed by MASafe.

Distinguished from other methods that attack default victim agents or select victim agents in ways that are not effective enough, the superior performance of AdapAM is due to its ability to learn victim-agent and malicious action in different environments and states.
It is worth mentioning that although MASafe demonstrates outstanding attack performance, this is because it is the only method that applies perturbations to all agents.
This approach leads to poor stealthiness, which is analyzed later.

\begin{table}[t]
    \centering
    \resizebox{\columnwidth}{!}{
    \begin{tabular}{c||c|c|c|c}
        \toprule
        \textbf{Env} & \textbf{AdapAM} & \textbf{MASafe} & \textbf{AMCA} & \textbf{Lin}
        \\ \midrule
        SMAC-1c3s5z & \textbf{0.14} & 0.35 & 0.25 & 0.19                
        \\ \midrule
        SMAC-8m & \textbf{0.12} & 0.33 & 0.21 & 0.15                    
        \\ \midrule
        SMAC-bane\_vs\_bane & \textbf{0.16} & 0.37 & 0.27 & 0.22        
        \\ \midrule
        SMAC-27m\_vs\_30m & \textbf{0.19} & 0.39 & 0.29 & 0.23         
        \\ \midrule
        GF-counterattack & \textbf{0.12} & 0.32 & 0.21 & 0.16          
        \\ \midrule
        GF-3\_vs\_1 & \textbf{0.10} & 0.31 & 0.20 & 0.15                
        \\ \midrule
        MPE-spread & \textbf{0.13} & 0.32 & 0.26 & 0.21                 
        \\ \midrule
        MPE-reference & \textbf{0.14} & 0.40 & 0.22 & 0.19             
        \\ \bottomrule
    \end{tabular}
    }
    \caption{Stealthiness: the magnitude of perturbations (the lower, the better), on the normal target MAS.}
    \label{tab:distance}
\end{table}

\begin{table}[t]
    \centering
    \resizebox{\columnwidth}{!}{
    \begin{tabular}{c||c|c|c|c|c}
        \toprule
        \textbf{Env} & \textbf{AdapAM } & \textbf{MASafe} & \textbf{AMCA} & \textbf{AMI} & \textbf{Lin}
        \\ \midrule
        SMAC-1c3s5z & \textbf{0.57} & 0.82 & 0.61 & 0.72 & 0.66             
        \\ \midrule
        SMAC-8m & \textbf{0.59} & 0.86 & 0.64 & 0.77 & 0.69                 
        \\ \midrule
        SMAC-bane\_vs\_bane & \textbf{0.39} & 0.90 & 0.58 & 0.67  & 0.61    
        \\ \midrule
        SMAC-27m\_vs\_30m & \textbf{0.36} & 0.91 & 0.57 & 0.65  & 0.62      
        \\ \midrule
        GF-counterattack & \textbf{0.61} & 0.87 & 0.68 & 0.76 & 0.71        
        \\ \midrule
        GF-3\_vs\_1 & \textbf{0.67} & 0.91 & 0.70 & 0.79 & 0.74             
        \\ \midrule
        MPE-spread & \textbf{0.63} & 0.85 & 0.69 & 0.78 & 0.74              
        \\ \midrule
        MPE-reference & \textbf{0.71} & 0.92 & 0.76 & 0.81 & 0.79           
        \\ \bottomrule
    \end{tabular}
    }
    \caption{Stealthiness: the F1 score when facing attack detection (the lower, the better), on the normal target MAS.}
    \label{tab:detection}
\end{table}

\subsection{Stealthiness Evaluation}
\label{sec:Stealthiness}
We evaluate the stealthiness from two aspects.
First, we measured the magnitude of perturbations added by different methods, calculating the distance between the perturbed observation $\tilde{o}$ and the original observation ${o}$ using the L-$\infty$ norm \cite{L-inf} as shown below:
\begin{equation}
\label{eq-L-inf}
    \| \tilde{o} - o\|_\infty = \max((\tilde{o}^{1} - o^{1}), \ldots, ({\tilde{o}^{m} - o^{m})}),
\end{equation}
where $\tilde{o}^x$ or $o^x$ represents the $x$-th element in the observation vector.
The smaller the L-$\infty$ distance, the more imperceptible the added perturbation is, i.e., better stealthiness.
Since AMI attack directly manipulates the action of agent rather than adding perturbations, we did not calculate its perturbation distance. Its stealthiness is evaluated solely based on the abnormal action detection, described below.
Table \ref{tab:distance} represents the results of the magnitude of perturbations.

In addition to the perturbation magnitude, we use the method proposed in \cite{detect} to detect the abnormal actions of victim agents.
This method predicts the action distribution of all agents based on the state and detects abnormal actions by calculating the normality scores.
The F1 score of the detection results is used to represent stealthiness. The lower the F1 score, the harder it is to detect, and the more stealthy it is.
Table \ref{tab:detection} lists the results.
% The results on the \textit{Robust Target MAS} are shown in Appendix \ref{sec:app-Results}.

\noindent\textbf{Results.}
As shown in Table \ref{tab:distance}, the perturbations added by AdapAM are the smallest in magnitude, outperforming the baselines. 
Moreover, the results in Table \ref{tab:detection} demonstrate that AdapAM is the most difficult to detect, outperforming the baselines.
Although MASafe exhibits an attack performance nearly comparable to AdapAM, and even outperforms AdapAM in one case, the results in Tables \ref{tab:distance} and \ref{tab:detection} indicate that the stealthiness of MASafe is the worst.
The reason is that the high attack performance of MASafe stems from the fact that the perturbations are applied to all agents, in a way that does not take stealthiness into account at all.

On the contrary, AdapAM guarantees stealthiness by adaptively choosing only one victim agent through the adaptive selection policy.
Besides, AdapAM leverages proxy agents, enabling the use of white-box-based adversarial example generation methods to easily craft undetectable perturbations, thereby ensuring its stealthiness.

\section{Conclusion}
This paper proposes AdapAM for effective and stealthy black-box adversarial attacks on MAS. AdapAM includes two key components: (1) an adaptive selection policy that adaptively selects victim agents and determines malicious actions to consider simultaneously effectiveness and stealthiness; (2) a perturbation generation module using proxy agents trained to approximate the target MAS, effectively generating observation perturbations injected to victims for executing malicious actions. Experimental results across eight environments demonstrate the superior attack performance and better stealthiness of AdapAM, compared to four baselines. 
Our future work will focus on extending AdapAM to larger and more diverse scenarios and designing robust defense MAS based on the insights from AdapAM.

%%%%%%%% Acknowledgments %%%%%%%%
\section{Acknowledgments}
This work was supported by the National Natural Science Foundation of China Grant No.62232016 and No.62506355, Basic Research Program of ISCAS Grant No. ISCAS-JCZD-202304 and No.ISCAS-JCZD-202405, Strategic Priority Research Program of Chinese Academy of Sciences Grant No. XDB0900000, Major Program of ISCAS Grant No. ISCAS-ZD-202302,
the National Research Foundation, Singapore, and the Cyber Security Agency under its National Cybersecurity R\&D Programme (NCRP25-P04-TAICeN). Any opinions, findings and conclusions or recommendations expressed in this material are those of the author(s) and do not reflect the views of National Research Foundation, Singapore and Cyber Security Agency of Singapore.
The authors would like to thank the anonymous reviewers for their valuable comments.

\bibliography{aaai2026}

\end{document}